\documentclass[reqno, eucal]{amsart}
\usepackage{amsmath,amssymb,amsthm,amscd}
\usepackage[mathscr]{eucal}

\numberwithin{equation}{section}

\newtheorem{theorem}{Theorem}

\newtheorem{proposition}[theorem]{Proposition}
\newtheorem{corollary}[theorem]{Corollary}

\theoremstyle{definition}

\newtheorem{example}[theorem]{Example}
\newtheorem{remark}[theorem]{Remark}

\newcommand{\Z}{{\mathbb Z}}

\newcommand{\Q}{{\mathbb Q}}
\newcommand{\Rm}{{\mathscr R}}

\begin{document}
\title[Tetrahedron equation and $R$ matrix for infinite dimensional modules]
{Tetrahedron equation and quantum $R$ matrices \\
for infinite dimensional modules of  $U_q(A^{(1)}_1)$ and $U_q(A^{(2)}_2)$}

\author{Atsuo Kuniba}
\email{atsuo@gokutan.c.u-tokyo.ac.jp}
\address{Institute of Physics, Graduate School of Arts and Sciences,
University of Tokyo, Komaba, Tokyo 153-8902, Japan}

\author{Masato Okado}
\email{okado@sci.osaka-cu.ac.jp}
\address{Department of Mathematics, Osaka City University, 
3-3-138, Sugimoto, Sumiyoshi-ku, Osaka, 558-8585, Japan}

\maketitle

\vspace{1cm}
\begin{center}{\bf Abstract}\end{center}
From the $q$-oscillator solution to the tetrahedron equation
associated with a quantized coordinate ring,
we construct solutions to the Yang-Baxter equation
by applying a reduction procedure formulated earlier by 
S. Sergeev and the first author.
The results are identified with the quantum $R$ matrices for the
infinite dimensional modules of $U_q(A^{(1)}_1)$ and 
$U_q(A^{(2)}_2)$ corresponding to an affinization 
of Verma modules of 
their subalgebras isomorphic to $U_q(sl_2)$ and $U_{q^4}(sl_2)$.

\vspace{0.2cm}

\section{Introduction}\label{sec:1}
The tetrahedron equation \cite{Zam80}
is a 3 dimensional (3d) extension of
the Yang-Baxter equation \cite{Bax}.
Among its many formulations, 
the homogeneous version of vertex type  
is a quartic equation on the matrix $\Rm$
acting on the tensor cube of a vector space $F$
representing the ``single spin" states.  
See (\ref{TE}).
The tetrahedron equation possesses the structure that 
if one of the components in $\Rm\in \mathrm{End}(F^{\otimes 3})$  
is evaluated away appropriately, 
the resulting object in $\mathrm{End}(F^{\otimes 2})$
satisfies the Yang-Baxter equation.
The eliminated space, let us call it the 3rd component, 
serves as the internal degrees of freedom 
of the local spins in the resulting 2d vertex model.
This reduction works naturally also for the compositions of 
$\Rm$'s in the 3rd component, which implies that 
a single solution to the tetrahedron equation generates 
an infinite sequence of solutions to the Yang-Baxter equation.
They correspond to the optional size in the 3rd direction
of the 3d lattice on which an integrable vertex model associated with 
$\Rm$ is defined.
This kind of connection between the 2d and 3d integrable systems 
has been studied from various viewpoints since \cite{BS0} for example.
In particular, it has been shown recently that the reduction scheme reproduces 
the conventional quantum $R$ matrices for some reducible  representations 
of $U_q(A^{(1)}_n)$ \cite{BS} and 
the spin representations of 
$U_q(B^{(1)}_n), U_q(D^{(1)}_n), U_q(D^{(2)}_{n+1})$ \cite{KS}.
In the former the reduction of the 3rd space is due to the trace,
whereas in the latter it is done by taking the matrix elements 
between special ``boundary vectors".
It is an open problem to clarify the grand picture of such  
2d-3d connections. 
See Section \ref{sec:gen}  for a unified formulation of the problem
under which the preceding results \cite{BS,KS} 
are allocated and explained  more precisely.

In this paper we add a further solution of the problem 
as a modest step toward a thorough understanding of the 
2d-3d connection.
We apply the reduction scheme based on the special boundary vectors \cite{KS}
to the solution $\Rm$ of the tetrahedron equation studied in 
\cite{KV,BS,BMS,S,KO1}\footnote{
See Section \ref{sec:2} for more accounts on the origin of the solution.}.
It acts on the tensor cube of the $q$-oscillator Fock space 
$F = \bigoplus_{m\ge 0}\Q(q)|m\rangle$.
It is the same $\Rm$ as the one used in the trace construction in \cite{BS}.
For simplicity we concentrate on the simplest ``single-site" situation
along the 3rd direction.
Then the resulting solution to the Yang-Baxter equation is the 
linear operator acting on $F\otimes F$, which defines 
an integrable 2d vertex model accommodating 
the $q$-oscillator Fock states on each edge.
Besides its explicit formula (Proposition \ref{pr:rf}), our main result is 
the identification with the quantum $R$ matrices for certain
infinite dimensional representations of $U_q(A^{(1)}_1)$ and $U_q(A^{(2)}_2)$ 
under some specialization (Theorem \ref{th:main}).
These representations are constructed from Verma modules of the subalgebras 
isomorphic to $U_q(sl_2)$ and $U_{q^4}(sl_2)$, respectively.
We find it natural to encounter 
the rank 1 quantum affine algebras since 
our construction corresponds to the single-site situation.
Nonetheless being able to capture the both of them is gratifying.
It stems from the choices of the two kinds of boundary vectors 
to which our construction can be applied.
A similar feature has been observed in the earlier work 
\cite[Remark 7.2]{KS} where 
the quantum $R$ matrices 
for $B^{(1)}_n, D^{(1)}_n$ and $D^{(2)}_{n+1}$ have been covered
from a single 3d $L$ operator.

The layout of the paper is as follows.
In Section \ref{sec:2} we recall the 
reduction procedure based on the boundary vectors \cite{KS} along the 
simplest one-site situation.
An explicit formula of the new solution to the Yang-Baxter equation
is given in Proposition \ref{pr:rf}.
In Section \ref{sec:3} we give the infinite dimensional representations of 
$U_q(A^{(1)}_1)$ and $U_q(A^{(2)}_2)$ relevant to our issue.
The quantum $R$ matrices for them are described in terms of 
spectral decompositions.
Section \ref{sec:main} contains the main result of the paper 
giving the precise relation between 
the solutions to the Yang-Baxter equation obtained in 
Sections \ref{sec:2} and \ref{sec:3}.
In Section \ref{sec:gen} we formulate 
a general class of problems 
that include the relevant results of this paper and \cite{BS,KS}
as a future direction of researches.
Some further results are announced.
A full solution of them will bring an important insight 
into the 2d-3d connection in integrable systems in a broad sense.

In this paper we shall explain the basic construction
and only state the main results.
The detail of the derivation and 
generalization to more general setting 
(see Remark \ref{re:n} and Section \ref{sec:gen})
will be given elsewhere. 
Throughout the paper we assume that $q$ is generic and 
use the following notations:
\begin{align*}
&(z;q)_m = \prod_{j=1}^m(1-z q^{j-1}),\;\;
(q)_m = (q; q)_m,\;\;
\binom{m}{n}_{\!\!q}= \frac{(q)_m}{(q)_n(q)_{m-n}} \;
(0 \text{ unless } 0 \le n \le m),\\
&[m]=[m]_q = \frac{q^m-q^{-m}}{q-q^{-1}}, \;\;
[m]_q! = \prod_{j=1}^m [j]_q,\;\;
\{x\} = \frac{x-x^{-1}}{q-q^{-1}},\;\;
(n)_+ = \max(n,0).
\end{align*}

\section{Reducing the tetrahedron equation to the Yang-Baxter equation}\label{sec:2}
Let  $F$ be a vector space and 
$\Rm \in \mathrm{End}(F^{\otimes 3})$.
Consider the tetrahedron equation, which is 
an equality in $\mathrm{End}(F^{\otimes 6})$:
\begin{align}\label{TE}
\Rm_{124}\Rm_{135}\Rm_{236}\Rm_{456}=
\Rm_{456}\Rm_{236}\Rm_{135}\Rm_{124},
\end{align}
where $\Rm_{ijk}$ acts as $\Rm$ on the 
$i,j,k$ th components from the left in the 
tensor product $F^{\otimes 6}$.

We recall the prescription in \cite{KS}  that reduces 
a solution to the tetrahedron equation to the one for the Yang-Baxter equation.
We restrict ourselves to the simplest ``single-site" situation relevant to the present paper.
See \cite{KS} for a more general treatment.

Suppose there are vectors
\begin{equation}\label{cv}
|\chi_s(x,y)\rangle
=|\chi_s(x)\rangle \otimes |\chi_s(xy)\rangle
\otimes |\chi_s(y)\rangle
\in F\otimes F\otimes F,
\end{equation}
where $x,y$ are extra (spectral) parameters such that
\begin{equation}\label{RX}
\Rm |\chi_s(x,y)\rangle   = |\chi_s(x,y)\rangle.
\end{equation}
The index $s$ is a label of possibly more than one such vectors.
Suppose also similar vectors exist in the dual space:
\begin{equation}\label{cvd}
\langle \chi_s(x,y)|
=\langle \chi_s(x)| \otimes
\langle \chi_s(xy)| \otimes
\langle \chi_s(y)|
\in F^*\otimes F^*\otimes F^*,
\end{equation}
with the property
\begin{equation}\label{XR}
\langle \chi_s(x,y) |\Rm =\langle \chi_s(x,y) | .
\end{equation}
Then evaluating the tetrahedron equation (\ref{TE}) between
$\langle \chi_s(x,y)|$ and
$|\chi_t(1,1)\rangle $\footnote{In general $|\chi_t(x',y')\rangle $
can be used. However, in our examples
treated later such a freedom is absorbed into elsewhere.}
on the 4,5,6 th components,
one gets the Yang-Baxter equation
\begin{equation}\label{YBE1}
\Rm_{12}(x)
\Rm_{13}(xy)
\Rm_{23}(y)
=
\Rm_{23}(y)
\Rm_{13}(xy)
\Rm_{12}(x)\;
\in \mathrm{End}(F \otimes F \otimes F),
\end{equation}
where indices again signify the non trivially acting components in 
$F^{\otimes 3}$, and 
\begin{align}\label{evR}
\Rm_{12}(z) = \langle \chi_s(z)| \Rm_{123} |\chi_t(1)\rangle
\in \mathrm{End}(F\otimes F) \otimes 1
\end{align}
for example.
Here the bracket means the evaluation with respect to the 3rd component.
Denoting (\ref{evR}) just by $\Rm(z)\in \mathrm{End}(F\otimes F)$, 
it is also convenient to introduce 
\begin{equation}\label{vrho}
\check{\Rm}(z) = \varrho(z) P \,\Rm(z),
\end{equation}
where $P(u \otimes v) =v \otimes u$
is the linear operator exchanging the components and 
$\varrho(z)$ is an arbitrary scalar function.
Then the Yang-Baxter equation takes another familiar form:
\begin{equation}
\label{a:ybe}
(\check{\mathscr R}(x) \otimes 1)
(1\otimes \check{\mathscr R}(xy))
(1\otimes \check{\mathscr R}(y))
=(1\otimes \check{\mathscr R}(y))
(\check{\mathscr R}(xy) \otimes 1)
(1\otimes \check{\mathscr R}(x)).
\end{equation}
Note the degree of freedom to choose $s$ and $t$ in (\ref{evR}) 
although it has temporarily been suppressed in the notation.
In fact it will allow us to cover the 
quantum affine algebras for $A^{(1)}_1$ and $A^{(2)}_2$
in our main Theorem \ref{th:main}.

Now we proceed to a concrete realization of the above scheme in this paper.
We will always take $F$ to be an 
infinite dimensional space 
$F = \bigoplus_{m\ge 0}\Q(q)|m\rangle$.
The dual space will be denoted by 
$F^\ast = \bigoplus_{m\ge 0}\Q(q)\langle m|$
with the bilinear pairing 
$\langle m |n\rangle = (q^2)_m\delta_{m,n}$\footnote{
The dual space $F^\ast$ and this pairing will only be used in this section and 
Section \ref{sec:gen}.}.
For simplicity vectors like   
$|i\rangle \otimes|j\rangle \otimes|k\rangle \in F^{\otimes 3}$ and 
$\langle i | \otimes\langle j | \in (F^\ast)^{\otimes 2}$ etc. 
will be abbreviated to $|i,j,k\rangle$ and $\langle i,j|$ etc.

The solution $\Rm$ of the tetrahedron equation 
we are concerned with is the one 
obtained as the intertwiner of the quantum coordinate ring 
$A_q(sl_3)$ \cite{KV}\footnote{
The formula for it on p194 in \cite{KV} contains a misprint unfortunately.
Eq. (\ref{Rex}) here is a correction of it.},  
which was also found 
from a quantum geometry consideration in a different gauge including 
square roots \cite{BS, BMS}. 
They were shown to be essentially the same object 
and to constitute the solution of the 3d reflection equation \cite{KO1}.
It can also be identified with the transition matrix
of the PBW bases of the nilpotent subalgebra of $U_q(sl_3)$ \cite{S,KOY}. 
Here we simply call it 3d $\Rm$. 
It is given by
\begin{align}
\Rm |i,j,k\rangle &= 
\sum_{a,b,c} \Rm^{a,b,c}_{i,j,k}|a,b,c\rangle,\label{Rabc}\\
\Rm^{a,b,c}_{i,j,k} &=\delta^{a+b}_{i+j}\delta^{b+c}_{j+k}
\sum_{\lambda+\mu=b}(-1)^\lambda
q^{i(c-j)+(k+1)\lambda+\mu(\mu-k)}
\frac{(q^2)_{c+\mu}}{(q^2)_c}
\binom{i}{\mu}_{\!\!q^2}
\binom{j}{\lambda}_{\!\!q^2},\label{Rex}
\end{align}
where $\delta^m_n=\delta_{m,n}$ just to save the space.
The sum (\ref{Rex}) is over $\lambda, \mu \ge 0$ 
satisfying $\lambda+\mu=b$, which is also bounded by the 
condition $\mu\le i$ and $\lambda \le j$.
The formula (\ref{Rex}) is taken from \cite[eq.(2.20)]{KO1}, where
a proof of the property 
$(q^2)_a(q^2)_b(q^2)_c\,\Rm^{a,b,c}_{i,j,k} 
= (q^2)_i(q^2)_j(q^2)_k\,\Rm^{i,j,k}_{a,b,c}$ was also included.
Note that the elements of 3d $\Rm$ are polynomials in $q$ alone,
and the spectral parameter $z$ comes into the game only through 
the reduction (\ref{evR}).
Thus the notations $\Rm(z)$ and $\check{\Rm}(z)$ 
automatically distinguish them from the 3d $\Rm$, and they should be 
understood as the solutions to the Yang-Baxter equation.
We specify their matrix elements by
\begin{align}\label{RR}
\Rm(z)|i,j\rangle=\sum_{a,b}\Rm(z)^{a,b}_{i,j}|a,b\rangle,\quad
\check{\Rm}(z)|i,j\rangle=\sum_{a,b}\check{\Rm}(z)^{a,b}_{i,j}|a,b\rangle,
\quad \check{\Rm}(z)^{a,b}_{i,j} = \varrho(z){\Rm}(z)^{b,a}_{i,j}.
\end{align}

Let us turn to the vectors $|\chi_s(x,y)\rangle$ and 
$\langle \chi_s(x,y)|$ in (\ref{cv})--(\ref{XR}).
We use two such vectors obtained in \cite{KS}.
In the present notation they read
\begin{align}
&|\chi_1(z)\rangle = \sum_{m\ge 0}\frac{z^m}{(q)_m}|m\rangle,\quad
|\chi_2(z)\rangle = \sum_{m\ge 0}\frac{z^m}{(q^4)_m}|2m\rangle,
\label{xk}\\
&\langle \chi_1(z)| = \sum_{m\ge 0}\frac{z^m}{(q)_m}\langle m|,\quad
\langle \chi_2(z)| = \sum_{m\ge 0}\frac{z^m}{(q^4)_m}\langle 2m|.
\label{xb}
\end{align}

We define the four solutions to the Yang-Baxter equation
$\Rm(z) = \Rm^{s,t}(z)= \Rm^{s,t}(z,q)\, (s,t=1,2)$ by the formula
(\ref{evR}) in which (\ref{Rex}), (\ref{xk}) and (\ref{xb})
are substituted.
They are the matrices acting on $F \otimes F$ whose elements are given by
\begin{align}\label{rst0}
\Rm^{s,t}(z)^{a,b}_{i,j} &= \sum_{c,k \ge 0}
\frac{z^c(q^2)_{sc}}{(q^{s^2})_c(q^{t^2})_k}
\Rm^{a,b,sc}_{i,j,tk}.
\end{align}
Due to (\ref{Rex}) this is zero unless $a+b=i+j$ and 
the sum is actually a single one due to the constraint
$b+sc=j+tk$. 
In follows that $\Rm^{2,2}$ is decomposed as 
\begin{align}
\Rm^{2,2}(z) &= \Rm^{+,+}(z) \oplus \Rm^{+,-}(z) 
\oplus \Rm^{-,+}(z) \oplus \Rm^{-,-}(z),
\label{22pm}\\
\Rm^{\epsilon_1, \epsilon_2}(z) &\in 
\mathrm{End}(F^{\epsilon_1}\otimes F^{\epsilon_2}),\quad
F^{\pm} = \bigoplus_{m\ge0, (-1)^m=\pm1}\Q(q)|m\rangle.\label{fpm}
\end{align}
It implies $\check{\Rm}^{\epsilon_1, \epsilon_2}(z):
F^{\epsilon_1}\otimes F^{\epsilon_2}\rightarrow 
F^{\epsilon_2}\otimes F^{\epsilon_1}$.
For example $\check{\Rm}^{+,-}(z)$
is just the submatrix of $(\check{\Rm}^{2,2}(z)^{a,b}_{i,j})$
with the indices $a,j$ restricted to be odd and $b,i$ to be even.
Another notable fact is 
\begin{equation}
\Rm^{2,1}(z)^{a,b}_{i,j} = 
\frac{(q^2)_i(q^2)_j}{(q^2)_a(q^2)_b}z^{\frac{j-b}{2}}\,
\Rm^{1,2}(z^{\frac{1}{2}})^{i,j}_{a,b},
\end{equation}
which can easily be derived from the property of 
$\Rm^{a,b,c}_{i,j,k}$ mentioned after (\ref{Rex}).
Henceforth we concentrate on 
$\Rm^{1,1}(z), \Rm^{2,2}(z)$ and $\Rm^{1,2}(z)$
in the rest of the paper.

\begin{remark}\label{re:n}
The prescription \cite{KS} applied to the $n$-site setting 
leads to the four solutions of the Yang-Baxter equation
$(s,t=1,2)$
acting on $F^{\otimes n} \otimes F^{\otimes n}$ 
whose elements are given by  
\begin{align*}
\Rm^{s,t}(z)^{{\bf a}, {\bf b}}_{\,{\bf i}, \,{\bf j}}
=\sum_{c_0, \ldots, c_n\ge 0} 
\frac{z^{c_n}(q^2)_{sc_n}}{(q^{s^2})_{c_n}(q^{t^2})_{c_0}}
\prod_{l=1}^n\Rm^{a_l, b_l, c'_l}_{i_l, j_l, c'_{l-1}},\qquad
c'_l = \begin{cases}
tc_0 &l=0,\\
sc_n &l=n,\\
c_l & \text{otherwise},
\end{cases}
\end{align*}
where ${\bf a}= (a_1,\ldots, a_n)$ etc. 
The quantity (\ref{rst0}) corresponds to the $n=1$ case.
In the more general problem formulated in Section \ref{sec:gen},
this corresponds to ${\mathscr R}^{s,t}(z|\varepsilon_1,\ldots, \varepsilon_n)$
(\ref{rst}) with $\varepsilon_1= \cdots = \varepsilon_n = 0$.
\end{remark}

By a direct calculation we have 
\begin{proposition}\label{pr:rf}
For $(s,t) \in \{(1,1), (2,2), (1,2)\}$, the following formula is valid:
\begin{align}
&\Rm^{s,t}(z)^{a,b}_{i,j}= \delta^{a+b}_{i+j}\;
\delta^{(-1)^{(s+1)b}}_{(-1)^{(s+1)j}}\; 
z^{\varepsilon+\frac{1}{s}(j-b-\varepsilon)_+}(1+\varepsilon q)
\sum_{m,n,\lambda, \mu, \lambda+\mu=b}
(-1)^{(\kappa-1)(t-1)m+n+\lambda}q^\phi
\nonumber \\
&\qquad\times 
\binom{\frac{|b-j|-\varepsilon}{\kappa}}{m}_{\!\!q^{\kappa^2}}
\binom{\min(b,j)-\lambda}{n}_{\!\!q^2}
\binom{i}{\mu}_{\!\!q^2}
\binom{j}{\lambda}_{\!\!q^2}
\frac{((-1)^sz^{\frac{t}{s}}
q^{t(\lambda-\mu+i+\kappa m+2n+\varepsilon)+s}; q^{st})_\infty}
{(z^{\frac{t}{s}}q^{t(\lambda-\mu+i+\kappa m+2n)}; q^{st})_\infty},
\label{mass}\\
&\phi=\frac{m}{2}(\kappa^2m-\kappa+2)+n(n+2|b-j|+1)-i\min(b,j)
+\mu^2+\lambda+(\lambda-\mu)(b-j)_+\nonumber \\
&\qquad +\varepsilon(4m+2n+\lambda-\mu+i),\\
&\kappa = \begin{cases}
2 & {\rm if } \; (s,t)=(1,2),\; b \ge j,\\
s & {\rm otherwise},
\end{cases}\qquad
\varepsilon = \begin{cases}
1 & {\rm if } \; (s,t)=(1,2),\; b-j \in 2\Z_{\ge 0} +1,\\
0 & {\rm otherwise}.
\end{cases}
\end{align}
The sum in (\ref{mass}) is over $m,n,\lambda, \mu\ge 0$
with the constraint $\lambda + \mu=b$.
It is a finite sum due to the support property of the $q$-binomial 
coefficients. 
The second Kronecker delta postulates 
$b\equiv j$ mod 2 when $s=2$, which guarantees 
$(j-b-\varepsilon)_+/s, \,(|b-j|-\varepsilon)/\kappa \in \Z$.
\end{proposition}

Denoting the scalar function $\varrho(z)$ in (\ref{vrho}) and (\ref{RR}) 
for $\Rm^{s,t}(z)$ by $\varrho^{s,t}(z)$, we choose it as
\begin{align}
\varrho^{s,t}(z) = \left(
\frac{\bigl(z^{\frac{t}{s}}; q^{st}\bigr)_\infty}
{\bigl((-1)^sq^s z^{\frac{t}{s}}; q^{st}\bigr)_\infty}\right)^{\epsilon_1\epsilon_2},
\end{align}
where for $(s,t)=(2,2)$, the signs 
$\epsilon_1, \epsilon_2$ are to be taken according to 
the four components in (\ref{22pm}).
For $(s,t)=(1,1)$ and $(1,2)$, the quantity $\epsilon_1\epsilon_2$
is to be interpreted as $+1$. 
Then it follows from Proposition \ref{pr:rf} that 
the matrix elements of $\check{\Rm}^{s,t}(z)$ are 
rational functions of $q$ and $z$.
They are useful for computer checks of the 
Yang-Baxter equation (\ref{a:ybe}).

\begin{example}\label{ex:M}
Let $M^{s,t}_d=(\check{\Rm}^{s,t}(z)^{i,d-i}_{j,d-j})_{0 \le i,j \le d}$ be the 
matrix where $i$ and $j$ are the row and the column indices
taking $(0,0)$ at the top left corner.
One has $M^{s,t}_0=(1)$ for any $s,t$.
\begin{align*}
&M^{1,1}_1=\left(
\begin{array}{cc}
 \frac{(q+1) z}{q z+1} & \frac{1-z}{q z+1} \\
 \frac{q (z-1)}{q z+1} & \frac{q+1}{q z+1}
\end{array}
\right),
\;
M^{1,2}_1=\left(
\begin{array}{cc}
 \frac{(q+1) z}{q z^2+1} & \frac{1-z^2}{q z^2+1} \\
 \frac{q \left(z^2-1\right)}{q z^2+1} & \frac{(q+1) z}{q z^2+1}
\end{array}
\right),
\;
M^{2,2}_1=\left(
\begin{array}{cc}
 0 & \frac{1}{1-z} \\
 \frac{q}{z-1} & 0
\end{array}
\right),\\
&M^{1,1}_2=\left(
\begin{array}{ccc}
 \frac{(q+1) \left(q^2+1\right) z^2}{(q z+1) \left(z q^2+1\right)} 
& -\frac{(q+1) (z-1) z}{(q z+1) \left(z q^2+1\right)} &
   \frac{(z-1) (q z-1)}{(q z+1) \left(z q^2+1\right)} \\
 \frac{q (q+1) \left(q^2+1\right) (z-1) z}{(q z+1) \left(z q^2+1\right)} 
& \frac{z q^3+2 z q^2-q^2-z^2 q+2 z q+z}{(q z+1) \left(zq^2+1\right)} 
& -\frac{(q+1) \left(q^2+1\right) (z-1)}{(q z+1) \left(z q^2+1\right)} \\
 \frac{q^2 (z-1) (q z-1)}{(q z+1) \left(z q^2+1\right)} & 
 \frac{q (q+1) (z-1)}{(q z+1) \left(z q^2+1\right)} & \frac{(q+1)
   \left(q^2+1\right)}{(q z+1) \left(z q^2+1\right)}
\end{array}
\right),\\
&M^{1,2}_2 =
\left(
\begin{array}{ccc}
 \frac{(q+1) z^2 \left(z^2 q^3-z^2 q^2+q^2+1\right)}{\left(q z^2+1\right) \left(z^2 q^3+1\right)} & 
 -\frac{(q+1) (z-1) z
   (z+1)}{\left(q z^2+1\right) \left(z^2 q^3+1\right)} & 
   \frac{(z-1) (z+1) (q z-1) (q z+1)}{\left(q z^2+1\right) \left(z^2
   q^3+1\right)} \\
 \frac{q (q+1) \left(q^2+1\right) (z-1) z (z+1)}{\left(q z^2+1\right) \left(z^2 q^3+1\right)} & 
 \frac{z^2 q^4+z^2 q^3-z^4 q^2+2
   z^2 q^2-q^2+z^2 q+z^2}{\left(q z^2+1\right) \left(z^2 q^3+1\right)} & 
   -\frac{q (q+1) \left(q^2+1\right) (z-1) z (z+1)}{\left(q
   z^2+1\right) \left(z^2 q^3+1\right)} \\
 \frac{q^2 (z-1) (z+1) (q z-1) (q z+1)}{\left(q z^2+1\right) \left(z^2 q^3+1\right)} & 
 \frac{q^2 (q+1) (z-1) z (z+1)}{\left(q
   z^2+1\right) \left(z^2 q^3+1\right)} & 
   \frac{(q+1) \left(z^2 q^3+z^2 q-q+1\right)}{\left(q z^2+1\right) \left(z^2
   q^3+1\right)}
\end{array}
\right),\\
&M^{2,2}_2=
\left(
\begin{array}{ccc}
 \frac{\left(q^2-1\right) z}{q^2 z-1} & 0 & \frac{z-1}{q^2 z-1} \\
 0 & \frac{q^2-z}{q^2 z-1} & 0 \\
 \frac{q^2 (z-1)}{q^2 z-1} & 0 & \frac{q^2-1}{q^2 z-1}
\end{array}
\right),\\
&M^{2,2}_3=
\left(
\begin{array}{cccc}
 0 & \frac{z-q^2 z}{(z-1) \left(q^4 z-1\right)} & 0 & \frac{1-q^2 z}{(z-1) \left(q^4 z-1\right)} \\
 \frac{q \left(q^6-1\right) z}{(z-1) \left(q^4 z-1\right)} & 0 & \frac{q z-q^3}{(z-1) \left(q^4 z-1\right)} & 0 \\
 0 & \frac{q^2 \left(q^2-z\right)}{(z-1) \left(q^4 z-1\right)} & 0 & \frac{1-q^6}{(z-1) \left(q^4 z-1\right)} \\
 \frac{q^3 \left(q^2 z-1\right)}{(z-1) \left(q^4 z-1\right)} & 0 & \frac{q \left(q^2-1\right)}{(z-1) \left(q^4 z-1\right)} & 0
\end{array}
\right).
\end{align*}
\end{example}

\section{Quantum $R$ matrices for infinite dimensional modules}\label{sec:3}
The quantum affine algebras (without derivation operator) 
$U_q(A^{(1)}_1)$ and $U_q(A^{(2)}_2)$ are the Hopf algebras 
generated by $e_i, f_i, k^{\pm 1}_i\, (i=0,1)$ satisfying the relations
\begin{align*}
&k_i k^{-1}_i = k^{-1}_i k_i = 1,\quad [k_i, k_j]=0,\\
&k_ie_jk^{-1}_i = q_i^{a_{ij}}e_j,\quad 
k_if_jk^{-1}_i = q_i^{-a_{ij}}f_j,\quad
[e_i, f_j]=\delta_{ij}\frac{k_i-k^{-1}_i}{q_i-q^{-1}_i},\\
&\sum_{\nu=0}^{1-a_{ij}}(-1)^\nu
e^{(1-a_{ij}-\nu)}_i e_j e_i^{(\nu)}=0,
\quad
\sum_{\nu=0}^{1-a_{ij}}(-1)^\nu
f^{(1-a_{ij}-\nu)}_i f_j f_i^{(\nu)}=0\;\;(i\neq j),
\end{align*}
where $e^{(\nu)}_i = e^\nu_i/[\nu]_{q_i}!, \,
f^{(\nu)}_i = f^\nu_i/[\nu]_{q_i}!$, 
$(q_0,q_1) = (q,q)$ for $A^{(1)}_1$ and $(q^4,q)$ for $A^{(2)}_2$.
The $(a_{ij})_{0\le i,j \le 1}$ is the Cartan matrix:
\begin{align*}
(a_{ij})_{0\le i,j \le 1} 
= \begin{pmatrix}2 & -2\\ -2 & 2\end{pmatrix} 
\;\text{for } A^{(1)}_1,\quad
\begin{pmatrix}2 & -1\\ -4 & 2\end{pmatrix}
\;\text{for } A^{(2)}_2.
\end{align*}
We use the coproduct of the form
\begin{align*}
\Delta k^{\pm 1}_i = k^{\pm 1}_i\otimes k^{\pm 1}_i,\quad
\Delta e_i = 1\otimes e_i + e_i \otimes k_i,\quad
\Delta f_i = f_i\otimes 1 + k^{-1}_i\otimes f_i.
\end{align*}

Let us introduce a $U_q(A_1^{(1)})$-module structure on the space $F$.
\begin{proposition}
The following defines a $U_q(A^{(1)}_1)$-module structure on $F$.
\begin{equation}
\begin{split}
&e_1|m\rangle = -[m]\{\alpha q^{m-1}\}|m-1\rangle,\quad
f_1|m\rangle = |m+1\rangle,\quad
k_1|m\rangle = \alpha^{-1}q^{-2m}|m\rangle,\\
&e_0 = xf_1,\quad f_0 = x^{-1}e_1, \quad k^{\pm 1}_0 = k^{\mp 1}_1,
\end{split}
\end{equation}
where $\alpha$ and $x$ are nonzero parameters.
\end{proposition}
For generic $\alpha$ it is irreducible, which will be denoted by $V_x(\alpha)$.

\begin{remark}\label{re:sl21}
As a module over $U_q(sl_2)$ generated by $e_1,f_1,k_1$, 
the space $F$ is identified with 
a Verma module, namely, it is generated by $f_1$ from an eigenvector $|0\rangle$
of $k_1$ killed by $e_1$. Such a module has already been known in \cite{Kul}, although
eigenvalues are special there.
\end{remark}

\begin{proposition}
The following defines an irreducible $U_q(A^{(2)}_2)$-module structure on $F$.
\begin{equation}
\begin{split}
&e_0|m\rangle  = -\varepsilon_0\frac{[2m][2m-2]}{[4]^2}|m-2\rangle,\quad
f_0|m\rangle = |m+2\rangle,\quad
k_0|m\rangle  = \varepsilon_0q^{-4m-2}|m\rangle,\\
&e_1|m\rangle =x |m+1\rangle,\quad
f_1|m\rangle =  \varepsilon_1(-1)^mx^{-1}\frac{[2m]}{[2]}|m-1\rangle,\quad
k_1|m\rangle = \varepsilon_1(-1)^m q^{2m+1}|m\rangle
\end{split}
\end{equation}
where $\varepsilon_0^2=\varepsilon_1^2=1$ and $x$ is a nonzero parameter.
\end{proposition}

\begin{remark}\label{re:sl20}
As a module over $U_{q^4}(sl_2)$ generated by $e_0,f_0,k_0$, 
the space $F$ is decomposed into
two components $F^+$ and $F^-$ in (\ref{fpm}). 
Both are Verma modules but with different eigenvalues.
\end{remark}

In what follows we will be exclusively concerned with the case 
$\varepsilon_0=\varepsilon_1=1$, which will be denoted by $V_x$.
Note that we have assigned $x$ (the spectral parameter) to the color $1$
generators rather than color $0$ for $U_q(A^{(2)}_2)$.
As the vector spaces, $V_x(\alpha)$ and $V_x$ are the same as $F$.

The quantum $R$ matrix 
$\check{R}(z)=\check{R}(z,q|\alpha,\beta)$ 
for our modules of 
$U_q=U_q(A^{(1)}_1), U_q(A^{(2)}_2)$ is a linear operator 
\begin{align}
\check{R}(z): V_x(\alpha) \otimes V_y(\beta) \rightarrow V_y(\beta)\otimes V_x(\alpha)
\quad (z=x/y),
\end{align}
where $\alpha, \beta$ dependence should be dropped 
for $U_q(A^{(2)}_2$).
The $R$ matrix is characterized by the intertwining relations \cite{Ji}
\begin{align}\label{ir}
\Delta(g) \check{R}(z) = \check{R}(z) \Delta(g)  \quad (\forall g \in U_q)
\end{align}
and the normalization which we choose as
$\check{R}(z)|0,0\rangle = |0,0\rangle$.
Define its matrix elements by 
\begin{align}
\check{R}(z)|i,j\rangle = \sum_{a,b}\check{R}(z)^{a,b}_{i,j}|a,b\rangle,
\end{align}
where the sum is over those $a,b \ge 0$ satisfying 
$a+b=i+j$ due to the weight conservation 
implied by (\ref{ir}) with $g=k_0, k_1$.

We include a description of the quantum $R$ matrices 
in terms of their spectral decomposition.
For $U_q(A^{(1)}_1)$ we set
\begin{align}
v^{(d)}_{\alpha,\beta} = \sum_{j=0}^d
(-q^{1-d}\beta^{-1})^{j}\,
\binom{d}{j}_{\!\!q^2}\,\prod_{k=1}^j
\frac{\{\alpha q^{d-k}\}}{\{ \beta q^{k-1}\}}\,|d-j,j\rangle
\in V_x(\alpha)\otimes V_y(\beta).
\end{align}
Up to an overall constant, this vector is characterized by
the conditions
\begin{align}
(\Delta k_1)v^{(d)}_{\alpha,\beta} = 
(\alpha\beta)^{-1}q^{-2d}v^{(d)}_{\alpha,\beta},
\qquad
(\Delta e_1) v^{(d)}_{\alpha,\beta}=0.
\end{align}

\begin{proposition}[$U_q(A^{(1)}_1)$ case]
The following direct sum decomposition holds as the 
module over $U_q(sl_2)$ mentioned in Remark \ref{re:sl21}:
\begin{align}
V_x(\alpha)\otimes V_y(\beta) = \bigoplus_{d\ge 0} V^{(d)},\quad
V^{(d)}=\bigoplus_{r \ge 0} \Q(q)(\Delta f_1)^r v^{(d)}_{\alpha,\beta}.
\end{align}
On each subspace $V^{(d)}$, $\check{R}(z)=\check{R}(z,q |\alpha, \beta)$ 
acts diagonally as $(z=x/y)$
\begin{align}
\check{R}(z) (\Delta f_1)^r v^{(d)}_{\alpha,\beta} &= 
\sigma^{(d)}_{\alpha,\beta}(z)
(\Delta f_1)^r v^{(d)}_{\beta,\alpha},\quad
\sigma^{(d)}_{\alpha,\beta}(z) =
\Bigl(\frac{\beta z}{\alpha}\Bigr)^{\!d}\,
\frac{(\alpha^2; q^2)_d(\alpha\beta z^{-1};q^2)_d}
{(\beta^2; q^2)_d(\alpha\beta z;q^2)_d}.\label{sigab}
\end{align}
\end{proposition}
The $R$ matrix satisfies the Yang-Baxter equation
(common $q$-dependence is suppressed)
\begin{equation}\label{ybe}
\begin{split}
&(\check{R}(y/z|\beta,\gamma)\otimes 1)
(1\otimes \check{R}(x/z|\alpha,\gamma))
(\check{R}(x/y|\alpha, \beta)\otimes 1)\\
&=(1\otimes \check{R}(x/y|\alpha,\beta))
(\check{R}(x/z|\alpha,\gamma)\otimes 1)
(1\otimes \check{R}(y/z|\beta,\gamma)),
\end{split}
\end{equation}
which is an equality of the maps 
$V_{x}(\alpha)\otimes 
V_{y}(\beta)\otimes 
V_{z}(\gamma) \rightarrow
V_{z}(\gamma) \otimes V_{y}(\beta)\otimes V_{x}(\alpha)$.
The inversion relation 
$\check{\Rm}(z|\alpha,\beta)\check{\Rm}(z^{-1}|\beta,\alpha)
=\mathrm{Id}$ is valid.

For $U_q(A^{(2)}_2)$ we set
\begin{align}
u^{(d)}_{\pm} = \sum_{j=0}^d(-1)^{j(j\mp 1)/2}\;q^{j(j-1)}
\binom{d}{j}_{\!\!q^4}|j,d-j\rangle\;\;
\in  V_x\otimes V_y.
\end{align}
Thus $u^{(0)}_+ = u^{(0)}_-=|0,0\rangle$.
They are characterized by the conditions
\begin{align}
(\Delta k_0)u^{(d)}_\pm =q^{-4d-4}u^{(d)}_\pm,\quad
(\Delta e_0) u^{(d)}_\pm=0,\quad 
\check{R}(z)u^{(d)}_{\pm} = (\pm 1)^d\sigma^{(d)}(\pm z)u^{(d)}_{\pm},
\end{align}
where $\sigma^{(d)}(z)$ is specified in (\ref{sigd}) and $z=x/y$.

\begin{proposition}[$U_q(A^{(2)}_2)$ case]
The following direct sum decomposition holds as the 
module over $U_{q^4}(sl_2)$ mentioned in Remark \ref{re:sl20}
($V^{(0)}_+ = V^{(0)}_-$ is denoted by $V^{(0)}$):
\begin{align}
V_x\otimes V_y = 
V^{(0)} \oplus \bigoplus_{d\ge 1} 
(V^{(d)}_+\oplus V^{(d)}_-),\quad
V^{(d)}_{\pm}=\bigoplus_{r \ge 0} \Q(q)(\Delta f_0)^r u^{(d)}_{\pm},
\end{align}
On each subspace $V^{(d)}_{\pm}$, 
$\check{R}(z)=\check{R}(z,q)$ acts diagonally as
$(z=x/y)$
\begin{align}
\check{R}(z) (\Delta f_0)^r u^{(d)}_{\pm} &= (\pm1)^d\sigma^{(d)}(\pm z)
(\Delta f_0)^r u^{(d)}_{\pm},\quad
\sigma^{(d)}(z)= \prod_{m=1}^d
\frac{q^{2m-1}+(-1)^{m-1}z}{(-1)^{m-1}+q^{2m-1}z}.\label{sigd}
\end{align}
\end{proposition}
The $R$ matrix satisfies the Yang-Baxter equation (\ref{ybe})
without dependence on $\alpha,\beta,\gamma$.
The inversion relation $\check{\Rm}(z)\check{\Rm}(z^{-1})=\mathrm{Id}$
is valid.

\section{Main theorem}\label{sec:main}
In Section \ref{sec:2} we derived the solutions 
$\check{\Rm}^{s,t}(z)=\check{\Rm}^{s,t}(z,q)$ 
to the Yang-Baxter equation 
by a reduction of the 3d $\Rm$.
In Section \ref{sec:3} the quantum $R$ matrices 
of the rank 1 quantum affine algebras 
were determined in terms of their spectral decompositions.
The both of these matrices act on the infinite dimensional 
space $F\otimes F$.
Our main theorem presented below identifies them
up to a scalar multiple and a similarity transformation.
 
\begin{theorem}\label{th:main}
(i) The $\check{\Rm}^{1,1}(z)$ equals a similarity transformation of 
the $U_q(A^{(1)}_1)$  $R$ matrix 
$\check{R}(z,q|\alpha,\beta)$ specialized as
\begin{align}\label{aska}
\check{\Rm}^{1,1}(z,q^2)^{a,b}_{i,j} &=
(-iq)^{j-b}\check{R}(z,q| -iq, -iq)^{a,b}_{i,j},
\end{align}
where $i$ in $-iq$ means $\sqrt{-1}$ and is unrelated to the matrix indices.

\vspace{0.2cm}\noindent
(ii) The components $\check{\Rm}^{\pm,\pm}(z)$ 
and $\check{\Rm}^{\pm,\mp}(z)$ of $\check{\Rm}^{2,2}(z)$ 
are proportional to 
a similarity transformation of 
the $U_q(A^{(1)}_1)$ $R$ matrix 
$\check{R}(z,q|\alpha,\beta)$ specialized as
\begin{align}
&\check{\Rm}^{\epsilon_1,\epsilon_2}(z,q)^{a,b}_{i,j} =
r^{\epsilon_1, \epsilon_2} q^{\bar{j}-\bar{b}}\check{R}(z,q^2|
q^{2-\epsilon_1},q^{2-\epsilon_2})^{\bar{a},\bar{b}}_{\bar{i},\bar{j}}\quad
(\epsilon_1, \epsilon_2 = \pm 1),\label{yada}\\
&r^{+,+}=1, \;r^{+,-}=\frac{q}{z-1},\;
r^{-,+}=\frac{1}{1-z},\;
r^{-,-}=\frac{q^2-z}{q^2 z-1},
\end{align}
where $\bar{n}$ denotes the largest integer not exceeding $\frac{n}{2}$.

\vspace{0.2cm}\noindent
(iii) The $\check{\Rm}^{1,2}(z)$ equals a similarity transformation of 
the $U_q(A^{(2)}_2)$  $R$ matrix $\check{R}(z,q)$ as
\begin{align}
\check{\Rm}^{1,2}(z,-q^2)^{a,b}_{i,j} &=
q^{j-b}\check{R}(z,q)^{a,b}_{i,j}.
\end{align}
\end{theorem}

In (ii) of the theorem,  
the general formula for $r^{\epsilon_1, \epsilon_2}$ is  
$\check{\Rm}^{2,2}(z,q)^{-\overline{\epsilon_2},-\overline{\epsilon_1}}_{
-\overline{\epsilon_1},-\overline{\epsilon_2}}$, 
which can be found in Example \ref{ex:M}.
The factors of the form $p^{j-b}$ is attributed to the 
similarity transformation by the operator 
$1\otimes K_p$ with $K_p|m\rangle=p^m|m\rangle$ 
which does not spoil the Yang-Baxter equation.
Combined with Proposition \ref{pr:rf},
Theorem \ref{th:main} provides an explicit formula 
for the quantum $R$ matrices of our infinite dimensional modules 
of the rank 1 quantum affine algebras.

Let $M^{s,t}_d(q)$ be 
the $(d+1)\times (d+1)$ matrix $M^{s,t}_d$ introduced in Example \ref{ex:M}
exhibiting the $q$-dependence.
We close the section with a corollary of Theorem \ref{th:main} giving the 
eigenvalues of 
$M^{1,1}_d(q^2), M^{2,2}_{2d}(q)$ and $M^{1,2}_d(-q^2)$\footnote{
The eigenvalues of $M^{2,2}_d(q)$ with odd $d$ is not 
directly derivable from (\ref{sigab}) since the matrix consists of 
the sub-matrices 
$\check{\Rm}^{\epsilon_1, \epsilon_2}(z)$ with $\epsilon_1\epsilon_2=-1$
which correspond, due to (\ref{yada}),  to the 
situation $\alpha\neq \beta$ in (\ref{sigab}).}.

\begin{corollary}
(i) The eigenvalues of 
$M^{1,1}_d(q^2)$ are given by 
the specialization of $\sigma^{(j)}_{\alpha, \beta}(z)$ (\ref{sigab}) as
\begin{align}
\{1, \sigma^{(1)}_{-iq,-iq}(z), 
\ldots, \sigma^{(d)}_{-iq,-iq}(z)\}
,\quad
\sigma^{(j)}_{-iq,-iq}(z) = \prod_{m=1}^j
\frac{z+q^{2m}}{1+zq^{2m}}.
\end{align}

\vspace{0.2cm}\noindent
(ii) The eigenvalues of 
$M^{2,2}_{2d}(q)$ are given by 
the specialization of $\sigma^{(j)}_{\alpha, \beta}(z)$ (\ref{sigab}) as
\begin{align}
\{1, \tilde{\sigma}^{(1)}(z)^{\times 2},  
\ldots, \tilde{\sigma}^{(d)}(z)^{\times 2}\}
,\quad
\tilde{\sigma}^{(j)}(z) = 
\sigma^{(j)}_{\alpha,\beta}(z)|_{q\rightarrow q^2, \,\alpha=\beta=q}
=\prod_{m=1}^j
\frac{z-q^{4m-2}}{1-zq^{4m-2}},
\end{align}
where the superscript $\times 2$ stands for the two-fold degeneracy.

\vspace{0.2cm}\noindent
(iii) The eigenvalues of $M^{1,2}_d(-q^2)$ are given by 
$\sigma^{(j)}(z)$ (\ref{sigd}) as
\begin{equation}
\begin{split}
&\{
\pm \sigma^{(1)}(\pm z), \pm \sigma^{(3)}(\pm z), \ldots, 
\pm \sigma^{(d)}(\pm z)\}\quad \text{if $d$ is odd},\\
&\{1, \sigma^{(2)}(\pm z), \sigma^{(4)}(\pm z), \ldots, 
\sigma^{(d)}(\pm z)\}
\quad \text{if $d$ is even}.
\end{split}
\end{equation}
\end{corollary}
These results can be directly checked for small $d$ 
by using Example \ref{ex:M}.

\section{Generalizations}\label{sec:gen}
The result in this paper is regarded as the solution of a special case
of a more general problem, which we shall now explain.
First we recall the 3d $L$ operator \cite{BS} in the form adapted to the
present context.
Let $F = \bigoplus_{m\ge 0}\Q(q)|m\rangle$ be the Fock space 
and $F^\ast$ be its dual as before.
Set $V = \Q(q)v_0 \oplus \Q(q) v_1$.
By the 3d $L$ operator we mean the following: 
\begin{align}
{\mathscr L} &=({\mathscr L}_{\alpha, \beta}^{\gamma,\delta})
 \in \mathrm{End}(V \otimes V \otimes F),
\quad
{\mathscr L}(v_\alpha \otimes v_\beta \otimes |m\rangle)
= \sum_{\gamma,\delta}v_\gamma \otimes v_\delta \otimes 
{\mathscr L}_{\alpha, \beta}^{\gamma,\delta}|m\rangle,
\end{align}
where there are six nonzero ${\mathscr L}_{\alpha, \beta}^{\gamma,\delta}
\in \mathrm{End}(F)$ given by
\begin{align}
{\mathscr L}_{0, 0}^{0,0}&= {\mathscr L}_{1,1}^{1,1} = 1,\;\;
{\mathscr L}_{0, 1}^{0,1} = {\bf k},\;\;
{\mathscr L}_{1,0}^{1,0} = -q{\bf k},\;\;
{\mathscr L}_{1,0}^{0,1} = {\bf a}^-,\;\;
{\mathscr L}^{1,0}_{0,1} = {\bf a}^+.
\end{align}
The operators ${\bf a}^\pm, {\bf k} \in \mathrm{End}(F) $ 
are called $q$-oscillators and act on $F$ by
\begin{align}
{\bf a}^+|m\rangle = |m+1\rangle,\quad
{\bf a}^-|m\rangle = (1-q^{2m})|m-1\rangle,\quad
{\bf k}|m\rangle = q^m|m\rangle.\label{ac1}
\end{align}
In short, the 3d $L$ operator ${\mathscr L}$ represents 
a six-vertex model having the $q$-oscillator valued Boltzmann weights.
It satisfies the tetrahedron equation \cite{BS}:
 \begin{align}\label{RLLL}
{\mathscr R}_{1,2,3}{\mathscr L}_{b,c,3}{\mathscr L}_{a,c,2}{\mathscr L}_{a,b,1}=
{\mathscr L}_{a,b,1}{\mathscr L}_{a,c,2}{\mathscr L}_{b,c,3}{\mathscr R}_{1,2,3}.
\end{align}
This is an equality in 
$\mathrm{End}(\overset{a}{V}\otimes \overset{b}{V} \otimes \overset{c}{V}
\otimes \overset{1}{F} \otimes \overset{2}{F} \otimes \overset{2}{F})$,
where 
$\overset{a}{V},\overset{b}{V},\overset{c}{V}$ are the copies of $V$ and 
$\overset{1}{F},\overset{2}{F},\overset{3}{F}$ are the ones for $F$.
The indices of ${\mathscr R}$ and ${\mathscr L}$ signify the 
components of the tensor product on which these operators act non trivially.
Viewed as an equation on ${\mathscr R}$, 
(\ref{RLLL}) is equivalent  \cite{KO1} to the 
intertwining relation of the irreducible representations of the 
quantized coordinate ring $A_q(sl_3)$ \cite{KV} in the sense that
the both lead to the same solution given in (\ref{Rex}) 
up to an overall  normalization.

Next we introduce the notation allowing us to teat 
${\mathscr R}$ and ${\mathscr L}$ on an equal footing.
\begin{align}
W^{(0)}=F,\quad
W^{(1)}=V,
\quad
{\mathscr S}^{(0)}={\mathscr R},
\quad
{\mathscr S}^{(1)}={\mathscr L}.
\end{align}
The ${\mathscr S}^{(r)}$ acts on 
$W^{(r)} \otimes W^{(r)}  \otimes F$ for $r=0,1$.
In what follows 
the copy of the Fock space $F$ that  plays the role analogous to the 
auxiliary space of the transfer matrices will be designated by ``3".
Define ${\bf h}_3$ acting on it by 
${\bf h}_3 | m \rangle = m|m \rangle$.

Let $n$ be any positive integer.
Consider the copies of 
$W^{(\varepsilon_i)}$ 
denoted by $\overset{\alpha_i}{W}{}^{(\varepsilon_i)}$  
and 
$\overset{\beta_i}{W}{}^{(\varepsilon_i)}$ for
$i=1,2, \ldots, n$.
We write their tensor product as 
\begin{align}
\overset{\alpha}{\bf W}{}^{({\bf \varepsilon})} = 
\overset{\alpha_1}{W}{}^{(\varepsilon_1)}\otimes \cdots \otimes 
\overset{\alpha_n}{W}{}^{(\varepsilon_n)} \;\;
\text{for }\;
 {\bf \varepsilon} = 
(\varepsilon_1, \varepsilon_2, \ldots, \varepsilon_n),
\end{align}
and similarly for 
$\overset{\beta}{\bf W}{}^{({\bf \varepsilon})}$
and 
$\overset{\gamma}{\bf W}{}^{({\bf \varepsilon})}$.
The labels $\alpha, \beta$ and $\gamma$ of the copies are put just for distinction
and these spaces are the same as  
$W^{(\varepsilon_1)}\otimes \cdots \otimes W^{(\varepsilon_n)}$
as vector spaces.

We introduce the five families of 
$R$ matrices each consisting of the $2^n$ members labeled with 
$(\varepsilon_1, \varepsilon_2, \ldots, \varepsilon_n) \in \{0,1\}^n$
as follows:
\begin{align}
{\mathscr R}^{\mathrm{tr}}(z|\varepsilon_1, \varepsilon_2, \ldots, \varepsilon_n)
&= \mathrm{Tr}_3\bigl(z^{{\bf h}_3} 
{\mathscr S}^{(\varepsilon_1)}_{\alpha_1, \beta_1,3}\,
{\mathscr S}^{(\varepsilon_2)}_{\alpha_2, \beta_2,3}
 \cdots {\mathscr S}^{(\varepsilon_n)}_{\alpha_n, \beta_n,3}\bigr),
\label{rtr}\\
{\mathscr R}^{s,t}(z|\varepsilon_1, \varepsilon_2, \ldots, \varepsilon_n)
&=
\langle \chi_s(z) | 
{\mathscr S}^{(\varepsilon_1)}_{\alpha_1, \beta_1,3}\,
{\mathscr S}^{(\varepsilon_2)}_{\alpha_2, \beta_2,3}
 \cdots {\mathscr S}^{(\varepsilon_n)}_{\alpha_n, \beta_n,3}
| \chi_t(1) \rangle \quad (s,t=1,2).\label{rst}
\end{align}
Here the bracket is evaluated in the Fock space $3$ by 
$\langle m | m'\rangle = \delta_{m,m'}(q^2)_m$ and 
the trace is by $\mathrm{Tr}_3(X) = \sum_{m\ge 0}
\frac{\langle m | X | m \rangle}{(q^2)_m}$.
They are linear operators acting on 
$\overset{\alpha}{\bf W}{}^{({\bf \varepsilon})} 
\otimes \overset{\beta}{\bf W}{}^{({\bf \varepsilon})}$.
By the construction and the 
tetrahedron equations (\ref{TE}) and (\ref{RLLL}) 
together with the properties (\ref{RX}) and (\ref{XR}),
we have
\begin{theorem}\label{th:gen}
Denote any one of (\ref{rtr}) and (\ref{rst}) with a fixed 
$(\varepsilon_1, \varepsilon_2, \ldots, \varepsilon_n)$ by 
${\mathscr R}_{\alpha, \beta}(z)$.
Then it satisfies the Yang-Baxter equation in
$\mathrm{End}(
\overset{\alpha}{\bf W}{}^{({\bf \varepsilon})} \otimes
\overset{\beta}{\bf W}{}^{({\bf \varepsilon})} \otimes
\overset{\gamma}{\bf W}{}^{({\bf \varepsilon})})$: 
\begin{equation}\label{YBE2}
\Rm_{\alpha, \beta}(x)
\Rm_{\alpha, \gamma}(xy)
\Rm_{\beta, \gamma}(y)
=
\Rm_{\beta, \gamma}(y)
\Rm_{\alpha, \gamma}(xy)
\Rm_{\alpha, \beta}(x).
\end{equation}
\end{theorem}

These solutions ${\mathscr R}_{\alpha, \beta}(z)$ 
to the Yang-Baxter equation are rational functions of the 
parameter $q$ and  the spectral parameter $z$ 
up to an overall scalar function of $z$.
Thus it is natural to seek their origin in the 
conventional quantum group theory.
We formulate it as

\vspace{0.2cm}\noindent
{\bf Problem}.
Find the appropriate quantum affine algebra 
and its  (possibly infinite dimensional) representation
by which the ${\mathscr R}_{\alpha, \beta}(z)$ is 
characterized as the intertwiner of the tensor product 
up to a normalization.
So far it has been studied in the following cases.
(The two $q$'s entering ${\mathscr R}_{\alpha, \beta}(z)$
and $U_q$ are not necessarily the same.)  

\vspace{0.2cm}

\begin{enumerate}

\item 
${\mathscr R}^{\mathrm{tr}}(z|,0,0,\ldots,0)$
was claimed to be the direct sum
$\oplus_{J,J'\ge 0}R_{J\omega_1,J'\omega_1}$ 
of the $U_q(A^{(1)}_{n-1})$ quantum $R$ matrices for 
the symmetric representations (relative 
normalization of the summands left unspecified) \cite{BS}. 

\item 
${\mathscr R}^{s,t}(z|1,1, \ldots, 1)$ with $(s,t)=(2,1), (2,2) $ and $(1,1)$
were identified \cite{KS} with the quantum $R$ matrices for the 
spin representations of $U_q(B^{(1)}_n), U_q(D^{(1)}_n)$ \cite{O} and 
$U_q(D^{(2)}_{n+1})$, respectively.
It was also suggested \cite{KS} that the case $(s,t)=(1,2)$ is the quantum $R$ 
matrix of the spin representation of $U_q(B^{(1)}_n)$ 
corresponding to the realization of $B^{(1)}_n$ as an affinization 
of its classical subalgebra $D_n$ rather than the standard $B_n$.

\item
${\mathscr R}^{s,t}(z|0)$ with $(s,t)=(1,1), (2,2)$ (resp. $(1,2), (2,1)$) 
are identified with 
the $U_q(A^{(1)}_1)$ (resp. $U_q(A^{(2)}_2)$) 
quantum $R$ matrices for the infinite dimensional representations
corresponding to an affinization of the Verma module of their classical subalgebras
in this paper.

\end{enumerate}

\vspace{0.2cm}
The full solution of the problem is a feasible task and 
will shed a valuable insight into 
the relation between 2d and 3d integrable systems.
In fact the $s=t=1$ case of the above result (3) can be extended to the 
$n$-site situation.
We have proved that  
${\mathscr R}^{1,1}(z|0,0,\ldots,0)$ is the 
quantum $R$ matrix of $U_q(D^{(2)}_{n+1})$ 
associated with the affinization of the Verma ($q$-oscillator) module of its 
classical subalgebra $U_q(B_n)$. 
It is consistent with (\ref{aska})
in view of  $D^{(2)}_2=A^{(1)}_1$
and is also natural from  \cite[Remark 7.2]{KS}.

It is interesting to 
investigate the effect of the mixture of ${\mathscr R}$ and ${\mathscr L}$ 
in (\ref{rtr}) and (\ref{rst})
which is firstly formulated here explicitly.
We finish by presenting 
an explicit formula of ${\mathscr R}^\mathrm{tr}(z|0,1)$
as the simplest example.

Note the decomposition ${\bf W}^{(0,1)} = \bigoplus_{d\ge 0}W^{(d)}$,
where $W^{(0)}=\Q(q)|0\rangle \otimes v_0$ and 
$W^{(d)}=\Q(q)|d\rangle \otimes v_0 \oplus 
\Q(q)|d\!-\!1\rangle \otimes v_1$ for $d\ge 1$.
Accordingly ${\mathscr R}^\mathrm{tr}(z|0,1)$ splits into the 
direct sum of the matrices acting on 
$W^{(d)}\otimes W^{(d')}$.
It turns out that they are zero unless $d=d'=0$ or $dd'\ge 1$.
Thus ${\mathscr R}^\mathrm{tr}(z|0,1)$ consists of 
a $1\times 1$ matrix and infinitely many $4\times 4$ matrices
corresponding to $\mathrm{End}(W^{(d)}\otimes W^{(d')})$
with $dd'\ge 1$.
Explicitly it is expressed as 
\begin{align}
{\mathscr R}^\mathrm{tr}(z|0,1)
=\frac{1}{1-z}\mathrm{Id}_{0,0}
\oplus \bigoplus_{d,d'\ge 1}
\frac{z^{d'-1}(q^{d-d'+2}z^{-1};q^2)_{d'-1}}
{(q^{d-d'}z;q^2)_{d'+1}}
(1\otimes K^{-1}){\mathscr R}_{q^d,q^{d'}}(z)(K\otimes 1),
\end{align}
where $\mathrm{Id}_{0,0}$ denotes the $1\times 1$ identity matrix and 
$K(|m\rangle \otimes v_\alpha) = q^{1-2\alpha}|m\rangle \otimes v_\alpha$
causes just a gauge transformation. 
The ${\mathscr R}_{q^d,q^{d'}}(z)\in \mathrm{End}(W^{(d)}\otimes W^{(d')})$ 
is given by the specialization of
\begin{align}\label{Rmn}
{\mathscr R}_{\mu,\nu}(z) = 
\begin{pmatrix}
z-\mu\nu  & 0&0 &0 \\
0 & \nu-\mu z & (1-\nu^2)z & 0\\
0 & 1-\mu^2 & \mu - \nu z & 0\\
0 & 0 & 0 & 1-\mu \nu z
\end{pmatrix}.
\end{align} 
By this we mean the linear operator acting as 
$\xi_0\otimes \eta_0
\mapsto (z-\mu\nu) \xi_0\otimes \eta_0$,
$\xi_0\otimes \eta_1
\mapsto (\nu-\mu z)\xi_0\otimes \eta_1
+ (1-\mu^2)\xi_1\otimes \eta_0$, etc.
in terms of the basis $\xi_i\otimes \eta_j$ of 
$W^{(d)}\otimes W^{(d')}$ taken as
$\xi_i = |d-i\rangle \otimes v_i$ and 
$\eta_j = |d'-j\rangle \otimes v_j$ for $i,j=0,1$.
The case $\mu=\nu$ is known to be the 
intertwiner of the quantum affine super algebra 
$U_q(\widehat{sl}(1|1))$.
The ${\mathscr R}_{\mu,\nu}(z)$ satisfies the 
Yang-Baxter equation 
${\mathscr R}_{\lambda,\mu}(x)
{\mathscr R}_{\lambda,\nu}(xy)
{\mathscr R}_{\mu,\nu}(y)=
{\mathscr R}_{\mu,\nu}(y)
{\mathscr R}_{\lambda,\nu}(xy)
{\mathscr R}_{\lambda,\mu}(x)$.
Further results including the detailed derivation and the proof 
of this paper will appear elsewhere.

\section*{Acknowledgments}
The authors thank Yasuhiko Yamada for collaboration in the previous work,
Kailash C. Misra and Yoshihisa Saito for communications on literature
and Tatsuya Toyoda for a careful reading of the manuscript.
A.K. thanks Vladimir Bazhanov, Vladimir Mangazeev 
and Sergey Sergeev for kind interest during his stay in Canberra in March 2012.
Especially he was benefited from the collaboration 
with S. Sergeev in \cite{KS}.  
This work is supported by Grants-in-Aid for
Scientific Research No.~23340007, No.~24540203 and
No.~23654007 from JSPS.

\end{document}